\documentstyle[12pt]{article}
\def\ZZ{{\mathchoice {\hbox{$\sf\textstyle Z\kern-0.4em Z$}}
{\hbox{$\sf\textstyle Z\kern-0.4em Z$}}
{\hbox{$\sf\scriptstyle Z\kern-0.3em Z$}}
{\hbox{$\sf\scriptscriptstyle Z\kern-0.2em Z$}}}}

\newcommand{\be}{\begin{equation}}
\newcommand{\ee}{\end{equation}}

\def\fun#1#2{\lower3.6pt\vbox{\baselineskip0pt\lineskip.9pt
\ialign{$\mathsurround=0pt#1\hfil##\hfil$\crcr#2\crcr\sim\crcr}}}

\begin{document}
\thispagestyle{empty}
\noindent\hspace*{\fill}  FAU-TP3-01/3 \\
\noindent\hspace*{\fill}  hep-ph/yymmdd \\
\noindent\hspace*{\fill}  4 June 2001   \\
\begin{center}\begin{Large}\begin{bf}
Domain-like Structures in the QCD Vacuum,\\
Confinement and Chiral-Symmetry Breaking
\end{bf}\end{Large}\vspace{.75cm}
 \vspace{0.5cm}

Alex C. Kalloniatis \footnote{akalloni@physics.adelaide.edu.au}
\\
Special Research Centre for the Subatomic Structure of Matter \\
University of Adelaide \\
South Australia 5005, Australia
\\
and \\
Sergei N. Nedelko \footnote{nedelko@theorie3.physik.uni-erlangen.de} \\
Institut f\"ur Theoretische Physik III
Universit\"at Erlangen - N\"urnberg \\
Staudtstra{\ss}e 7
D-91058 Erlangen, Germany
\\ and Bogoliubov Laboratory of Theoretical Physics, JINR,
141980 Dubna, Russia
\end{center}
\vspace{1cm}\baselineskip=35pt

\date{\today}
\begin{abstract} \noindent
We discuss how the inclusion of singular gauge fields in the
partition function for QCD can lead to a domain-like
picture for the QCD vacuum by virtue of specific
conditions on quantum fluctuations at the singularities.
With a simplified model of
hyperspherical domain regions with interiors of constant field strength
we calculate  the basic parameters of
the QCD vacuum, the gluon condensate, topological susceptibility,
string constant and quark condensate, and briefly discuss
confinement of dynamical quarks and gluons.
\end{abstract}
\vspace{1cm}
\begin{center}
{\it Presented at the Workshop on Lepton Scattering,
Hadrons and QCD,}
\\ {\it Adelaide, 26 March - 6 April, 2001.}
\end{center}
\newpage\baselineskip=18pt

\section{Introduction}
Singular gauge fields are most probably unavoidable in nonabelian 
gauge theories~\cite{singer}.
In the last two decades great effort has gone into clarifying     
whether or not (and, if yes, then how) such fields 
provide for confinement and chiral symmetry breaking.

We address this question in  a model based on an 
assumption that  singularities in vector 
potentials are concentrated on hypersurfaces $\partial V_j$ 
in Euclidean space, in the vicinity of which
gauge fields can be divided into a sum of a singular pure gauge  
$S_{\mu}$ and a regular fluctuation part 
$Q_{\mu}$, and where a certain colour vector $n_j^a$ can be associated with
$\partial V_j$.  For such configurations
not to have infinite action\cite{lenz} the fluctuation fields charged with 
respect to $n_j$ must obey specific boundary conditions on $\partial V_j$. 
The interiors of these regions thus constitute ``domains'' $V_j$.
Gauge field modes neutral with respect to $n^a_j$ are not restricted
and provide for interactions between domains. 
In a given domain $V_j$ the effect of fluctuations in neighbouring regions
can be seen as an external gauge field $B_{j\mu}^a$ neutral with respect
to $n_j^a$. This enables an approximation  in which
different domains are assumed to be decoupled from each other
but, to compensate, a certain mean field is introduced in domain interiors.

Decomposing a general gauge field $A^j_\mu=S^{j}_\mu+Q_\mu^j$ 
and demanding finiteness of the classical action we come to the conditions  
\begin{eqnarray}
\label{bc}
\breve n_j Q^{(j)}_\mu = 0, \ 
\psi=-i\!\not\!\eta^j e^{i\alpha_j\gamma_5}\psi,
\
\bar\psi=\bar\psi i\!\not\!\eta^j e^{-i\alpha_j\gamma_5},
\ x\in\partial V_j,
\end{eqnarray} 
with the adjoint matrix $\breve n_j=T^an_j^a$ appearing in the condition 
for gluons, and a bag-like boundary condition arising for quarks,   
with a unit vector $\eta^j_\mu(x)=x_{\mu}/|x|$ normal to $\partial V_j$. 

To make the model analytically tractable we consider
spherical domains with fixed radius $R$ and approximate the 
mean field in $V_j$ by a covariantly constant (anti-)self-dual 
configuration with the field strength
\begin{eqnarray}
&&\hat {\cal B}_{\mu\nu}^{(j)a}
=\hat n^{(j)}B^{(j)}_{\mu\nu},
\
\tilde B^{(j)}_{\mu\nu}=\pm B^{(j)}_{\mu\nu},
\
B^{(j)}_{\mu\nu}B^{(j)}_{\rho\nu}=B^2\delta_{\mu\rho},
\
B={\rm const},
\nonumber
\\
&&\hat n^{(j)}=t^3\cos\xi_j+t^8\sin\xi_j
\ (\xi_j\in\{\frac{\pi}{6}(2k+1),\ k=0,\dots,5\}),
\nonumber
\end{eqnarray}
where the parameter $B$ is the same for all domains
and the constant matrix $n_j^at^a$ belongs to the Cartan subalgebra.
It should be stressed that there is no source for this field on the boundary
and therefore it should be treated as strictly homogeneous in all further
calculations. The homogeneity itself appears here just as an approximation.

Thus the partition function we will deal with can be written 
\begin{eqnarray}
{\cal Z}&=&{\cal N}\lim_{V,N\to\infty}
\prod\limits_{i=1}^N\int_V\frac{d^4z_i}{V}
\int\limits_{\Sigma}d\sigma_i
\int_{{\cal F}^i_Q} {\cal D}Q^i\int_{{\cal F}^i_\psi}{\cal D}\psi_i 
{\cal D}\bar \psi_i
\nonumber\\
&{}&
\delta[D(\breve{\cal B}^{(i)})Q^{(i)}]
\Delta_{\rm FP}[\breve{\cal B}^{(i)},Q^{(i)}]
\exp\left\{
- S_{V_i}^{\rm QCD}
\left[Q^{(i)}+{\cal B}^{(i)},\psi^{(i)},\bar\psi^{(i)}\right]
\right\},
\nonumber
\end{eqnarray}
where the thermodynamic limit assumes $V,N\to\infty$ but 
with the density $v^{-1}=N/V$ taken fixed and finite. 
The measure of integration over parameters characterising domains is defined as
\begin{eqnarray}
\label{measure}
\int\limits_{\Sigma}d\sigma_i\dots&=&\frac{1}{48\pi^2}
\int\limits_0^{2\pi}d\alpha_i
\int\limits_0^{2\pi}d\varphi_i\int_0^\pi d\theta_i\sin\theta_i
\nonumber\\
&\times&\int_0^\pi d\omega_i\sum\limits_{k=0,1}\delta(\omega_i-\pi k)
\int_0^{2\pi} d\xi_i\sum\limits_{l=0,1,2}^{3,4,5}\delta(\xi_i-(2l+1)\pi/6)
\dots .
\end{eqnarray}
Here $\varphi_i$ and $\theta_i$ are spherical angles of
the chromomagnetic field,
$\omega_i$ is the angle between the chromomagnetic and
chromoelectric fields, $\xi_i$ is the angle in the colour
matrix $\hat n_i$, 
$\alpha_i$ is the chiral angle
and $z_i$ is the centre of the domain $V_i$.

This partition function describes a statistical system 
of density $N/V$ composed of noninteracting extended
domain-like structures, each of which is
characterised by a set of internal parameters and
whose internal dynamics are represented by the fluctuation fields. 

\section{Mean Field Correlators}  
In this model the connected $n-$point correlator 
\begin{eqnarray}
&&\langle B^{a_1}_{\mu_1\nu_1}(x_1)\dots B^{a_n}_{\mu_n\nu_n}(x_n) \rangle
= B^{n} t^{a_1\dots a_n}_{\mu_1\nu_1,\dots,\mu_n\nu_n}
\Xi_n(x_1,\dots,x_n),
\nonumber\\
&&t^{a_1\dots a_n}_{\mu_1\nu_1,\dots,\mu_n\nu_n}=
\int d\sigma_j
n^{(j)a_1}\dots n^{(j)a_n}B^{(j)}_{\mu_1\nu_1}\dots B^{(j)}_{\mu_n\nu_n},
\nonumber
\end{eqnarray}
of field strength tensors,
$$
 B^{a}_{\mu\nu}(x)
=\sum_j^N n^{(j)a}B^{(j)}_{\mu\nu}\theta(1-(x-z_j)^2/R^2),
$$
can be calculated explicitly using the measure, Eq.~(\ref{measure}). 
Translation-invariant functions
$$
\Xi_n(x_1,\dots,x_n)=\frac{1}{v}\int d^4z
\theta(1-(x_1-z)^2/R^2)\dots
\theta(1-(x_n-z)^2/R^2)
$$  
emerge and can be seen as the volume of the region of overlap of $n$ 
hyperspheres of radius $R$ and centres ($x_1,\dots,x_n$),
normalised to the volume of a single hypersphere
$v=\pi^2R^4/2$.
The functions $\Xi_n$ are continuous  and 
vanish if $|x_i-x_j|\ge 2R$. 
Correlations in the background field have finite range
$2R$. The Fourier transform of $\Xi_n$ is then an entire analytical 
function and thus correlations do not have a particle interpretation.
The statistical ensemble 
of background fields is not Gaussian since all connected correlators
are independent of each other and cannot be reduced 
to the two-point correlations. 

The simplest application of the above correlators gives  
a gluon condensate density which to this approximation is 
$$g^2 \langle F^a_{\mu\nu}(x)F^a_{\mu\nu}(x)\rangle=4B^2.$$

Another vacuum parameter which plays
a significant role in the resolution of the $U_A(1)$ problem
is the topological susceptibility 
\cite{Cre77}.
To define this we consider first the topological charge density
for the colour group $SU(3)$ in the lowest approximation 
\begin{eqnarray}
Q(x)={{g^2}\over {32 \pi^2}} \tilde F^a_{\mu\nu}(x)F^a_{\mu\nu}(x)
={{B^2}\over {8 \pi^2}} \sum_{j=1}^N\theta[1-(x-z_j)^2/R^2]\cos\omega_j,
\nonumber
\end{eqnarray} 
where $\omega_j\in\{0,\pi\}$ depends on the duality of the $j$-th domain.
For a given field configuration the topological charge is additive 
$$
Q=\int_V d^4x Q(x)= q(N_+ - N_-), \   q=B^2R^4/16, \  -Nq\le Q\le Nq
$$ 
where $q$ is a `unit' topological charge,
namely the absolute value of the topological charge of a single
domain, and $N_{+}$ $(N_-)$ is the number of 
domains with (anti-)self-dual field, $N=N_+ + N_-$. 
The probability of finding the topological charge $Q$ in a given 
configuration is defined by the distribution
$$
{\cal P}_N(Q)=\frac{{\cal N}_N(Q)}{{\cal N}_N}=
\frac{N!}{2^N\left(N/2-Q/2q\right)!\left(N/2+Q/2q\right)!},
$$
where ${\cal N}_N(Q)$ is the number of configurations with a given charge
and ${\cal N}_N$ is the total number of configurations.
The distribution is symmetric about $Q=0$, where it has a maximum
for $N$ even. 
For $N$ odd the maximum is at $Q=\pm q$. 
Averaged topological charge is zero. 

The topological susceptibility
$\chi$
is determined by the two-point correlator of topological charge density,
which in the lowest approximation gives 
\begin{equation}
\label{topsusc}
\chi = \int d^4x \langle Q(x) Q(0) \rangle=
{{B^4}\over {64 \pi^4}}
\int d^4x\Xi_2(x)
 = {{B^4 R^4} \over {128 \pi^2}}.
\nonumber
\end{equation}

\section{Area Law for the Wilson Loop}

To zeroth order in fluctuations the Wilson loop is given by the integral
\begin{eqnarray}
W(L)=\lim_{V,N\to\infty}\prod\limits_{j=1}^N\int_V\frac{d^4z_j}{V}
\int d\sigma_j\frac{1}{N_c}{\rm Tr}
\exp\left\{i\int_{S_L}d\sigma_{\mu\nu}(x)\hat B_{\mu\nu}(x)\right\}.
\nonumber
\end{eqnarray}
Note that path ordering in our case is not necessary since the matrices 
$\hat n^k$ are assumed to be in the Cartan subalgebra.
Computationally it is convenient to
consider a circular contour in the $(x_3,x_4)$
plane of radius $L$ with centre at the origin. 
To illustrate the steps in the calculation  we consider here 
the case of colour group 
$SU(2)$, 
though the final result for $SU(3)$ will be quoted below.
For colour $SU(2)$ we have $\hat n^k=\epsilon^k\tau_3, \ \epsilon^k=\pm 1.$
The thermodynamic limit  assumes that the density
$N/V$ is fixed. 
Calculation of the colour trace  gives
\begin{eqnarray}
&&W(L)=
\lim_{V,N\to\infty}\left[\int_V\frac{d^4z_j}{V}
\int d\sigma_j\frac{1}{2}
\left(e^{i B^j_{\mu\nu}J_{\mu\nu}(z_j)}
+
e^{-i B^j_{\mu\nu}J_{\mu\nu}(z_j)}
\right)\right]^N,
\nonumber\\
&&J_{\mu\nu}(z_k)=\int_{S_L}d\sigma_{\mu\nu}(x)\theta(1-(x-z_k)^2/R^2).
\nonumber
\end{eqnarray}
We have exploited the property that the integral above
does not depend on the index $j$.
As the contour of the Wilson loop is in the $(x_3,x_4)$-plane, the
only nonzero components of $J_{\mu\nu}$ are 
\begin{eqnarray}
&&J_{34}=-J_{43}=\int_{S_L}dx_3dx_4\theta(1-(x-z)^2/R^2),
\
B_{\mu\nu}J_{\mu\nu}=2BJ_{43}\cos\theta,
\nonumber
\end{eqnarray} 
where $\theta$ is the angle between chromoelectric field and the 
third coordinate axis. After integrating over 
the spatial orientations of the vacuum field the Wilson loop takes the form
\begin{eqnarray}
W(L)=\lim_{N,V\to\infty}\left[\frac{1}{V}\int_Vdz
\frac{\sin 2BJ_{43}(z)}{2BJ_{43}(z)}\right]^N.
\nonumber
\end{eqnarray} 
Integrating over $z$ and 
then taking the thermodynamic limit ($N\to\infty$, $v=V/N=\pi^2R^4/2$),
gives finally for a large Wilson loop $L\gg R$  
\begin{eqnarray}
&&W(L)=\lim_{N\to\infty}\left[1-\frac{1}{N}U(L)
\right]^N=e^{-U(L)}, \
U(L)=\sigma \pi L^2 + O(L), 
\nonumber\\
&&\sigma=Bf(BR^2), \ f(z)=\frac{2}{\pi z}
\left(1-\frac{1}{2\pi z}\int_0^{2\pi z}\frac{dx}{x}\sin x\right),
\nonumber
\end{eqnarray}
which displays an area law.
For $SU(3)$  the only difference is in the
function $f(z)$ which turns out to be: 
\begin{eqnarray}
\label{sig-su3}
f(z)=\frac{2}{3\pi z}
\left(3-
\frac{\sqrt{3}}{2\pi z}\int_0^{2\pi z/\sqrt{3}}\frac{dx}{x}\sin x
-
\frac{2\sqrt{3}}{\pi z}\int_0^{\pi z/\sqrt{3}}\frac{dx}{x}\sin x
\right).
\nonumber
\end{eqnarray}
The function $f$ is positive for $z>0$ and has a maximum for $z= 1.55$.
We choose this maximum to estimate the model parameters 
by fitting the string constant to the lattice result,
\begin{eqnarray}
\label{par-val}
\sqrt{B}=947{\rm MeV}, \ R^{-1}=760 {\rm MeV},
\end{eqnarray}
and get for the gluonic parameters of the vacuum
\begin{eqnarray}
\label{res1}
\sqrt{\sigma}=420 {\rm MeV}, \ \chi=(197 {\rm MeV})^4,
\ \frac{\alpha_s}{\pi}\langle F^2\rangle=0.081({\rm GeV})^4, \ q=0.15.
\end{eqnarray} 
If there were clear separation of the two scales characterising
the system (namely if either $\sqrt{B}R\gg 1$ or $\sqrt{B}R\ll 1$)
then an approximate treatment
of Green's functions for the fluctuation fields would be possible.
We observe that there is no separation: $\sqrt{B}R\approx 1$. 
Thus an approximation based on large or small domains is
not justifiable. 

Observe also that if $B$ goes to zero then the string constant vanishes.
This underscores the role of the gluon condensate in the confinement of 
static charges.  On the other hand, if the number of domains is fixed and 
the thermodynamic limit is 
defined as $V,R\to\infty, N={\rm const.}<\infty$, namely if  
the domains are macroscopically large,
then $W(L)=1$, which indicates the absence of a linear potential
between infinitely heavy charges in a purely homogeneous field. 

Since we have exactly integrated over background fields the 
role of finite range of correlation functions is hidden in the above 
calculation. In order to see this role explicitly one would need to decompose
the integrand into an infinite series and integrate term by term.
At this step all correlation functions of the 
background field up to infinite order would be manifest. The arguments 
about the importance of a fast decay of correlators
for confinement of static charges~\cite{DoSim} would be seen to apply here via
this representation.

\section{Confinement of Fluctuation Fields}

Confinement of the fluctuational fields can be studied via the  
analytical properties of their propagators. 
For the types of fields and boundary conditions
we have described here, the propagators of fluctuations
can be analytically calculated by reduction of the problems for both
quarks and gluons to that of a charged scalar field. 
The scalar case is essentially equivalent to the problem 
of a four-dimensional harmonic oscillator with the 
orbital momentum coupled to the external field, and
the general solution for these Green's functions 
can be found exactly by standard decomposition over hyperspherical harmonics.

Qualitatively it is clear that due to the boundary conditions we are dealing
with, the $x-$space propagators of charged fields are defined in
regions of finite support where they have  integrable
singularities so that their Fourier transforms 
are entire functions in the complex momentum plane.
This can be treated as confinement of dynamical fields\cite{leutw}. 
A known consequence of entire propagators is a Regge spectrum of 
relativistic bound states\cite{efi}, which is physically appealing.
A more expansive analysis of confinement in this
model will be given elsewhere\cite{NKYM01}.
 
\section{Quark Condensate}

Due to averaging over self- and anti-self-dual configurations and 
all possible values of the angle $\alpha$ in the partition function 
chiral symmetry is not broken explicitly.
However, as we show below, a nonzero quark condensate arises 
in the massless limit due to an interplay between the random distribution of 
domains with self- and anti-self-dual field and the boundary conditions
with random value of the chirality violating angle $\alpha$.

The quark propagator is defined by the equations
\begin{eqnarray}
&&(i\!\not\!\partial
-\frac{1}{2}\hat n \gamma_\mu B_{\mu\nu}x_\nu-m)S(x,y)=-\delta(x,y),
\\
&&i\!\not\!\eta(x)e^{i\alpha\gamma_5}S(x,y)=-S(x,y), \ (x-z)^2=R^2,
\nonumber\\
&&S(x,y)i\!\not\!\eta(y)e^{-i\alpha\gamma_5}=S(x,y), \ (y-z)^2=R^2,
\nonumber
\end{eqnarray}
where $\eta_\mu(x)=(x-z)_\mu/|x-z|$. 
Substitution
\begin{eqnarray}
&&S(x,y)=(i\!\not\!D+m)
[P_\pm{\cal H}_0+P_\mp O_+{\cal H}_1 + P_\mp O_-{\cal H}_{-1}]
\label{q-pr1},
\nonumber\\
 &&O_+=N_+\Sigma_+ + N_-\Sigma_-, \ O_-=N_+\Sigma_- + N_-\Sigma_+,
\nonumber\\ 
&&
N_\pm=\frac{1}{2}(1\pm \hat n/|\hat n|), \
\Sigma_\pm=\frac{1}{2}(1\pm \vec\Sigma\vec B/B), \ \hat B=|\hat n|B,
\nonumber
\end{eqnarray}
shows that scalar functions
 ${\cal H}_{\zeta}(x,y)$ should solve the equations:
\begin{eqnarray}
(-D^2+m^2+2 \zeta \hat B){\cal H}_{\zeta}=\delta(x,y),
\nonumber
\end{eqnarray}
If we were to look for solutions vanishing at infinity then the Green's 
function ${\cal H}_{-1}$ ($\zeta=-1$) would be divergent in the massless limit 
due to zero modes of the Dirac operator. The bag-like boundary conditions 
remove zero eigenvalues from the spectrum, and the massless limit is regular.

In order to avoid cumbersome calculations and render the
role of the former zero modes transparent, we turn to the particular 
choice $y=z=0$ and calculate the value of the quark condensate at the centre
of the domain. 
In this case  ${\cal H}_{\zeta}$ 
are functions of $x^2$ only, and the general solutions 
for scalar Green's functions take the form ($\mu_{\zeta}=m^2/2B + \zeta$)
\begin{eqnarray}
&&{\cal H}_{\zeta}=\Delta(x^2|\mu_{\zeta})+
C_{\zeta}\Phi(x^2|\mu_{\zeta}),
\ \Phi(x^2|\mu)=e^{-Bx^2/4}M(1+\mu,2,Bx^2/2).
\nonumber
\end{eqnarray}
Here $\Delta(x^2|\mu)$ is scalar propagator which vanishes at infinity 
with mass $2B\mu$, and $\Phi_\zeta$ is a solution to the homogeneous equation
regular at $x^2=0$,  
expressed in terms of the confluent hypergeometric function.
The constants $C_\zeta$ can be fit to implement the boundary condition.
The terms  $m{\cal H}_0$ and $m{\cal H}_1$ vanish in the massless limit
and do not contribute to the condensate.
The nontrivial contribution comes from $m{\cal H}_{-1}$.
The bag-like conditions imply that on the boundary
${\cal H}_{-1}$ satisfies a mixed condition, where $f'=df/d|x|$
and the sign ($-$)$+$  corresponds to an (anti-)self-dual domain,
\begin{eqnarray}
&&2e^{\mp i\alpha}m {\cal H}_{-1} = -2{\cal H}_{-1}' 
- \hat BR^2{\cal H}_{-1}.
\nonumber
\end{eqnarray} 
which leads to the relations
\begin{eqnarray}
&&\lim_{m\to 0}m{\cal H}_{-1}=
\frac{e^{\pm i\alpha}}{2\pi^2R^3}F(\hat BR^2/2)e^{-\hat Bx^2/4},
\
{\rm Tr}S(0,0)=\frac{e^{\pm i\alpha}}{2\pi^2R^3}\sum\limits_{|\hat n|}
F(\hat BR^2/2),
\nonumber\\
&&F(z)=e^z-z-1+\frac{z^2}{4}\int_0^\infty
\frac{dte^{2t-z({\rm coth}t-1)/2}}{{\rm sinh}^2t} 
({\rm coth}t-1).
\nonumber
\end{eqnarray}
Note that the term in the propagator with nonzero trace 
is proportional to the zero mode
of the Dirac operator: $\!\not\!DP_\mp O_{-}\exp(-\hat Bx^2/4)=0$.

Averaging this result over the angle $\alpha$ and 
(anti-)self-dual configurations and taking into 
account the $\alpha$-dependence of quark determinant~\cite{wipf} 
$
{\rm det}S^{-1}\propto \exp\{\pm i q\alpha\},
$
with $q$ being unit topological charge, we get a finite result
\begin{eqnarray}
\langle \bar\psi \psi\rangle= 
-\frac{q}{2\pi^2R^3(1+q)}\sum\limits_{|\hat n|}
F(\hat BR^2/2).
\nonumber
\end{eqnarray}
For  $B$ and $R$  fixed by the string constant as in Eq.~(\ref{par-val})
this is equal to
$$
\langle \bar\psi \psi\rangle= -(228 {\rm MeV})^3,
$$
which indicates spontaneous breaking of chiral symmetry.
Final conclusions require complete calculation
 including averaging over positions of domains.

\section*{Acknowledgments}
ACK is supported by the Australian Research Council.
We are grateful to Frieder Lenz 
for fruitful discussions and critical comments,
as well as to Garii Efimov, Jan Pawlowski and 
 Lorenz von Smekal for valuable
communication.

\end{document}